\documentclass[useAMS,usenatbib]{mn2e}\usepackage{times}
\input epsf

\newcommand{\xmm} {{\sl XMM-Newton }}
\newcommand{\chandra} {{\sl Chandra }}
\newcommand{\rosat} {{\sl ROSAT }}

\title[4C~34.16]{The cool wake around 4C~34.16 as seen by \xmm}
\author[Sakelliou et al.]{I. Sakelliou$^{1,2}$,  
D.M. Acreman$^{1}$, 
M.J. Hardcastle$^{3}$, 
M.R. Merrifield$^{4}$,  
T.J. Ponman$^{1}$, \and
I.R. Stevens$^{1}$  \\
$^{1}$School of Physics \& Astronomy, University of Birmingham, Edgbaston, Birmingham B15 2TT \\
$^{2}$Max-Planck-Institute f\"{u}r Astronomie, K\"{o}nigstuhl, 17, D-69117, Heidelberg, Germany \\
$^{3}$School of Physics, Astronomy \& Mathematics, University of Hertfordshire, College Lane, Hatfield, Hertfordshire AL10 9AB \\ 
$^{4}$School of Physics \& Astronomy, University of Nottingham, University Park, Nottingham NG7 2RD \\
}
\begin{document}

\pagerange{\pageref{firstpage}--\pageref{lastpage}} \pubyear{2005}

\maketitle

\label{firstpage}

\begin{abstract}
We present \xmm observations of the wake-radiogalaxy system 4C~34.16,
which shows a cool and dense wake trailing behind 4C~34.16's host
galaxy. A comparison with numerical simulations is enlightening, as
they demonstrate that the wake is produced mainly by ram pressure
stripping during the galactic motion though the surrounding
cluster. The mass of the wake is a substantial fraction of the mass of
an elliptical galaxy's X-ray halo. This observational fact supports a
wake formation scenario similar to the one demonstrated numerically by
Acreman et al (2003): the host galaxy of 4C~34.16 has fallen into its
cluster, and is currently crossing its central regions. A substantial
fraction of its X-ray halo has been stripped by ram pressure, and
remains behind to form the galaxy wake.

\end{abstract}


\section{Introduction}

As a cluster galaxy moves through the intracluster medium (ICM), an
X-ray bright region trails behind it, revealing its direction of
motion. We now know that there are two physical processes that take
place as a galaxy moves through the ICM, and lead to the creation of
an overdense wake behind it: i) the accretion of ICM (Bondi-Hoyle
accretion; Sakelliou 2000), and ii) ram pressure stripping of the
galactic material (Stevens et al. 1999). Naturally, both processes
should take place at the same time (e.g., Acreman et al. 2003).
 
Depending on the properties of the galaxy and the cluster [e.g., the
galaxy velocity ($v_{\rm gal}$) and the temperature of the ICM
($T_{\rm ICM}$)], wakes behind galaxies may be visible at X-ray
wavelengths. An understanding and knowledge of their production
mechanisms, of the most favourable conditions for detectable wake
production, and of their properties would be very rewarding for two
main reasons. Firstly, wakes provide the only general means of finding
the direction of the motion of the galaxy in the plane of the sky,
thus contributing to the study of the dynamics of clusters (Merrifield
1998). This method of studying clusters has been demonstrated already
using both \rosat (Drake et al. 2000) and \chandra data (Acreman et
al. 2005). Secondly, wakes are manifestations of the galaxy/ICM
interactions, which modify the properties of both the cluster (e.g.,
by contributing to the metal enrichment of the ICM) and the galaxy
(e.g., by inducing star formation). Thus, they may provide clues to
the transformation and subsequent evolution of galaxies in clusters.

Our aim was to find the X-ray properties of wakes that had been known
from the epoch of \rosat. Figure~1 shows the \xmm image of a first
example: the wake associated with the radio galaxy
4C~34.16. Unfortunately, not many candidates were known from \rosat
observations, and we did not have accurate measurements of their
properties.  \rosat images and spectra of a few systems (M86,
Rangarajan et al. 1995; 4C~34.16, Sakelliou et al. 1996; NGC4472,
Irwin \& Sarazin 1996; NGC1404, Jones et al. 1997) gave the first
clues that wakes exist, and that they might be cooler than the
surrounding medium.  The new instrumentation on \chandra and \xmm has
provided us with some more evidence that wakes might be a common
phenomenon: in deep X-ray images of clusters' cores wake-like features
have been seen (e.g., Abell~1795 Fabian et al. 2001; Abell~133 Fujita
et al. 2002; Abell~2199 Johnstone et al. 2002). Temperature maps of
limited resolution also suggest that these filaments are cooler than
their surrounding media.  As other authors have also argued (e.g.,
Fabian et al. 2001), wakes must be due to the motion of the central
cluster galaxy relative to the ICM. The recent work on the \xmm data
of M86 has been very instructive, as it has showed that the `plume'
associated with that galaxy is metal rich (Finoguenov et
al. 2004). However, in all these cases there has been no independent
evidence (at other wavelengths, for example) which could confirm that
the galaxy is moving in the direction indicated by the wake.

\begin{figure}
\begin{center} 
\leavevmode
\epsfxsize 1.0\hsize
\epsffile{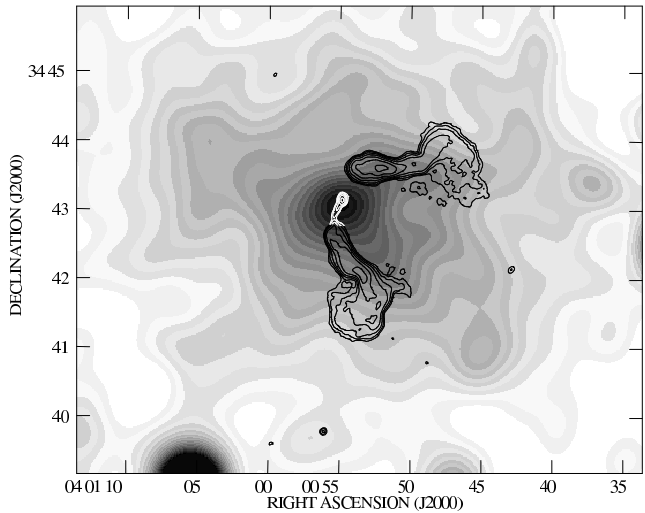}
\leavevmode
\caption{\xmm and radio data of the field around the radio galaxy
4C~34.16.  \xmm images from the three imaging detectors (MOS1, MOS2,
and PN) in the (0.3-5.0)~keV energy range have been co-added, and the
mosaic adaptively smoothed. Contours of the radio emission at 1.4~GHz
are overlaid.}\label{fig1}
\end{center} 
\end{figure}

Confirmation of the direction of motion of 4C~34.16 comes from the
shape of its radio jets. 4C~34.16 is a radio source associated with
the central galaxy in the cluster Z0357.9+9432. Its plumes are
symmetrically bent into a wide C-shape (see Fig.~\ref{fig1}). The most
widely favoured model invokes ram pressure as the main force
responsible for the plume bending.  Fortunately, the direction
indicated by the bent plumes coincides with the direction of the wake,
supporting the models of wake production discussed above.

In this paper we present \xmm observations (Section~2) of the
field around 4C~34.16. Using these data we derive the properties of
the cluster in Section~3, and of the wake in Section~4. Finally,
in Section~5 we discuss the most favourable models for the wake
production, and find the parameters that are consistent with the
observations.

Throughout this paper we use a redshift for 4C~34.16 of 0.078
[NASA/IPEC Extragalactic Database (NED)]. We adopt a Hubble constant
of $H_{0}=71 \ {\rm km \ s^{-1} \ Mpc^{-1}}$, and $\Omega_{\rm
M}=0.27$, which places the source at 349.3~Mpc, and gives a scale of
1.457 kpc arcsec$^{-1}$.

\section{Observations and Data reduction}

The field around 4C~34.16 was observed with \xmm for
$\sim$27~ksec. During the observation the two MOS and PN instruments
were operating in the PrimeFullWindow and PrimeFullWindowExtended mode
respectively; the medium filter was used for all three
instruments. The Optical Monitor (OM) was not switched on, due to the
presence of bright stars within the field of view.

The raw event lists from the EPIC instruments were processed and
calibrated with SAS v5.3. During the processing the
parameter {\it withbadpixfind} was switched on so that bad pixels that
had not been recorded in the calibration files were found and
subsequently removed. After the initial processing we confirmed that
new bad pixels were found. The calibrated events were filtered for
flags, using the \xmm flags {\sc $\#$XMMEA\_EM} and {\sc $\#$XMMEA\_EP}
for the two MOS and the PN detectors respectively. Restrictions on
the pattern were also applied: we kept only events with 
PATTERN$<$12 for the MOS cameras, and $<4$ for the PN.

The observation was contaminated by background flares, which were
apparent in all energy bands. To clean the event lists for image
analysis (Section~\ref{image}), the prescription in Read \& Ponman (2003) was
followed. This cleaning procedure left a useful exposure of
$\sim$21~ksec for each the two MOS cameras, and $\sim$13~ksec for the
PN camera. For the spectral analysis we imposed more strict limits,
which reduced the exposures further by $\sim$5~ksec (see Section~3.2).

Using the filtered and clean event files of each imaging \xmm
instrument, images and exposure maps were created in the (0.3-5.0)~keV
energy range. The images from all three instruments were finally
merged after being exposure corrected, using the SAS task {\sc
emosaic}. Subsequently, the mosaic was smoothed with an adaptive
kernel using {\sc CIAO}'s {\sc csmooth} tool.  Figure~\ref{fig1}
presents the merged image, with the contours of a 1.4-GHz radio map
overlaid.

The radio data consist of 1.4-GHz observations in the B and C
configurations of the NRAO Very Large Array (VLA). The B-configuration
data were taken by us on 2002 Aug 23; the C-configuration data were
obtained from the VLA data archive (programme AM79). Both datasets
were calibrated, merged and imaged within AIPS in the standard
manner.  As previously explained, the central cluster galaxy hosts
the double, bent radio source 4C~34.16. In Fig.~\ref{contours} we show
the contour plot of the merged (0.3-5.0)~keV image, overlaid onto the
DSS optical image of the field around the central galaxy in the
cluster, which is the host of 4C~34.16.

It is apparent from Fig.~\ref{fig1} and Fig.~\ref{contours} that the
cluster is fairly spherically symmetric on large scales. The core of
the radio galaxy coincides with the peak of the X-ray emission, which
is not located at the centre of the inner X-ray emission, but is
offset to the northeast.  The X-ray emission around the radio galaxy
is elongated along a line that bisects the angle between the two
radio lobes. As mentioned before, the shape of the radio galaxy
indicates a galactic motion to the North-East. Thus, the asymmetrical X-ray
emission could be due to a wake trailing behind the galaxy.

In the following sections we use the \xmm data to derive the
properties of the cluster (Section~3) and the wake (Section~4).

\section{Average Cluster Properties}\label{cluster}

\subsection{Image Analysis}\label{image}

For the purposes of the following spatial analysis, we generated
background subtracted and exposure corrected images independently for
each imaging instrument on \xmm. Firstly, images for all three
instruments were created in the (0.2-4.5)~keV energy range.  Following
Read \& Ponman (2003), particle and instrument background maps were
created. They were then subtracted from the data images, after scaling
them to the out-of-field events. The resulting images were corrected
for vignetting by multiplying them by the appropriate vignetting maps.

\begin{figure}
\begin{center} 
\leavevmode 
\epsfxsize 0.95\hsize
\epsffile{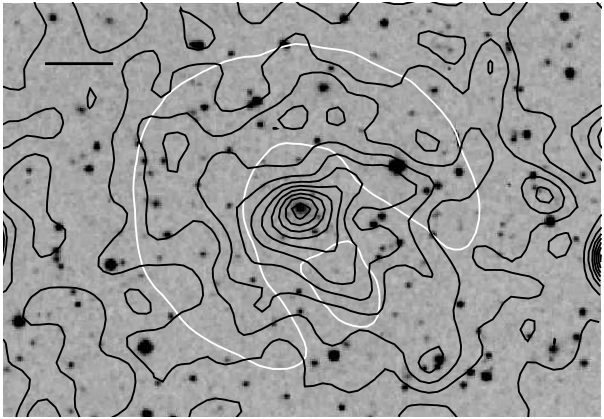}
\caption{X-ray contours of the adaptively smoothed (0.3-5.0keV) mosaic
overlaid onto the DSS image around 4C~34.16. The contour levels are
linearly spaced from (1-5)$\times 10^{-5} \ {\rm cnt \ s^{-1} \
pix^{-2}}$. The white contours mark the boundaries of the source
and background regions that are used for the spectral analysis of
Section~4.1. The horizontal bar at the top left corner of the image is
1~arcmin long.}\label{contours}
\end{center} 
\end{figure}

Our final aim was to compare the properties of the wake with those of its
environment, so that we can assess the wake production processes. Thus,
we used the above generated data images to derive the spatial
properties of the cluster. As Fig.~\ref{fig1}, and \ref{contours}
demonstrate, the X-ray emission from the wake and the galaxy dominate
in the inner cluster region, around the active core. At large
distances the cluster appears fairly symmetric.  In order to derive
the average cluster properties and compare them with past results, we
fitted the (0.2-4.5)~keV particle-subtracted
vignetting-corrected images in {\sc sherpa} with a 2-dimensional
$\beta$-model. Bright point sources were removed, by excluding circular
regions around them. We also decided to exclude the inner
$\sim$1.3~arcmin around the radio core, and we did not model the emission of
the X-ray halo and active nucleus during this fitting
procedure.  The 2-dimensional model was convolved with the
appropriate point-spread function (PSF) for each instrument, which is
a 2-dimensional image of the PSF of each instrument, as implemented
in the SAS task {\sc calview}.

Data from the three EPIC instruments were not co-added, but fitted
simultaneously. During the fitting procedure, the core radius ($r_{\rm
c}$) and $\beta$ parameter, of the $\beta$-model were linked and left
free to vary; the ellipticity was set to zero; and the centre of the
distribution was fixed to the location of the active core. The model
normalizations were allowed to vary independently for each
instrument. This fitting procedure resulted in a best-fitting value for
the core radius of $r_{\rm c}$
=2.371[2.336-2.412]~arcmin=207.26[204.23-210.89]~kpc, and a
$\beta$=0.858[0.829-0.890] (1$\sigma$ limits are quoted), consistent
with the \rosat results (Sakelliou et al. 1996). If the centre of the
$\beta$-model is left free to vary, the resulting parameters are
consistent within the errors with the ones we present above.

Although a derivation of the small scale galaxy's X-ray properties
(galactic X-ray halo and active nucleus) is outside the scope of this
paper, we attempted to model them. We added another $\beta$-model to
describe the galactic halo, and a delta function for the active
core. Again, the composite models were convolved with the appropriate
instrument PSFs. All trials reproduced the cluster properties found
before. However, the galaxy parameters were poorly constrained. Better
quality data are required to allow a determination of the galaxy properties.

\subsection{Spectral Analysis}\label{cluster_spec}

\begin{table}
\begin{center}
 \caption{Properties of the ICM and the Wake}\label{tab1}
\begin{tabular}{ccc}	\hline \hline

	&
ICM	&
WAKE	\\

\hline

$N_{\rm H}$	&
0.172[0.143-0.202]		&
0.689[0.288 - 1.280]		\\

($\times 10^{22} \ {\rm cm^{-2}}$)	&
					&
					\\

$T$		&
3.17[2.66-3.70]		&
1.14[0.39-1.94]		\\

(keV)			&
			&
			\\

$Z$		&
0.25[0.09-0.45]		&
0.25$^{f}$		\\

(${\rm Z_{\odot}}$)		&
			&
			\\

$\chi^{2}/d.o.f$	&
342.1/260			&
59.8/60		\\

\hline

$n$			&
1.33[1.02-1.64]		&
2.37[1.66-8.57]			\\

($\times 10^{-3} \ {\rm cm^{-3}}$)	&
			&
			\\

 \hline
\end{tabular}
\\
\vspace{0.2cm}
\begin{minipage}{8cm}
\small NOTES: All the limits reported are at the 90~per cent level, with $\Delta
\chi^2=2.71$. The abundance table is from Anders \& Grevesse (1989),
as implemented in {\sc xspec}.\\ 
$^{f}$: the parameter is frozen during the fitting procedure.
\end{minipage}
\end{center}
\end{table}

As mentioned in Section~2, for the purpose of spectral analysis
the original event lists were cleaned further leaving useful exposures
of $\sim$15.9, $\sim$16.0, and $\sim$10.2~ksec for the MOS1, MOS2, and PN
respectively.

To obtain a spectrum of the ICM, we accumulated counts in a circular
region centred on the central galaxy and extending out to
5~arcmin. The background spectrum was taken from the same observation,
in a annulus adjacent to the source region, and extending from 5 to
8.3~arcmin. Point-source regions were removed from the source and
background regions. The emission from the host galaxy of 4C~34.16 and
the wake were also excluded from the source region.  Here, as in any
subsequent spectral analysis, responses (rmfs) and auxiliary response
files (arfs) were generated with {\sc rmfgen-1.48.5} and {\sc
arfgen-1.54.7} respectively.

The spectrum in the (0.3-8.0)~keV energy range was modelled in {\sc
xspec} by a {\it mekal} thermal model modified by the line-of-sight
hydrogen column, as described by the {\sc xspec} {\sc wabs} model.
The Galactic column in the direction of 4C~34.16 is $N_{\rm H,G}=0.162
\times 10^{22} \ {\rm cm^{-2}}$. The hydrogen column density
($N_{\rm H}$), temperature of the ICM ($T_{\rm ICM}$), metal
abundances ($Z$), and normalization were left as free parameters. The
results of the spectral analysis of the cluster are tabulated in
Table~1, along with the spectral properties of the wake (these will be
derived in Section~\ref{wake_spec}).

Here it should be noted that the  $kT_{\rm ICM}$ adopted by Sakelliou
et al. (1996) was 1~keV. Their temperature was based on relatively
poor spectral fits to the {\it ROSAT} PSPC data. However, in their
analysis and interpretation they demonstrated how a larger $kT_{\rm
ICM}$ would have changed the final conclusions. In the following
sections it will be apparent how a more accurate measurement of the
cluster temperature affects the final results.

In the same table (Table 1) we present the number densities ($n$) of
both the cluster and the wake. The tabulated value for the ICM's
density is the central number density ($n_0$), derived from the
central surface brightness applying eq.~(4) from Sakelliou et
al. (1996). We calculated $n_0$ for each EPIC instrument independently
using the best fit values of the normalizations of the three
$\beta$-models found in Section~3.1. The value presented in Table~1 is
the weighted mean and error of these three values.

\section{The wake}

Galactic wakes appear as  excesses of X-ray
emission in the images of galaxies in clusters. Such an enhancement
 is seen in the (0.3-5.0)~keV image of 4C~34.16 (see
Fig.~\ref{contours}).

\begin{figure}
\begin{center} 
\leavevmode 
\epsfxsize 1.0\hsize
\epsffile{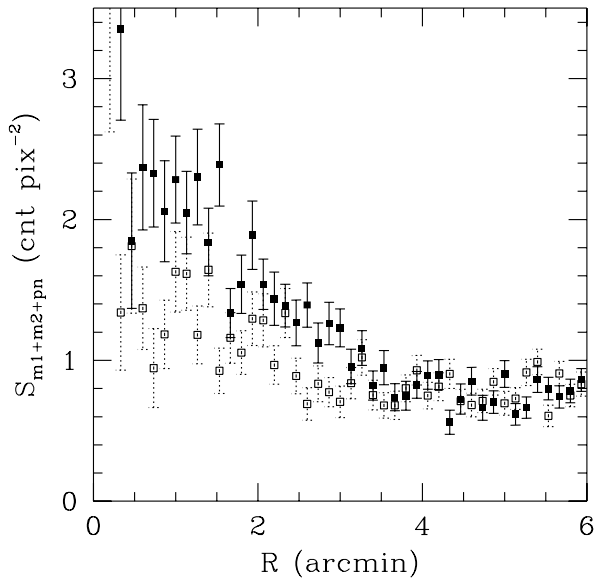}
\caption{Comparison of the radial profiles around 4C~34.16 in two
sectors. The `in-front' profile is shown by open squares, and the
`back' one by filled squares (see Section~4 for details).}\label{frontback}
\end{center} 
\end{figure}

The plot of Fig.~\ref{frontback} clearly shows the
asymmetry between the front and behind the galaxy. In this plot we
compare the radial profiles taken in regions that lie in front of and
behind the galaxy. For the derivation of these profiles we used the
background subtracted, (0.2-4.5)~keV, mosaic image discussed in
Section~3.1.  Both profiles were centred on the core of 4C~34.16.
Counts were accumulated in concentric annuli of 8~arcsec width. The
`in-front' profile was extracted in a sector between (100-140) degrees,
and the `back' one in a sector between (280-320)~degrees, where zero
degrees coincides with the positive $x$-axis of the image, and angles
are measured anti-clockwise.  This plot shows clearly that there is an
excess of X-ray emission behind the galaxy, and that the disruption
extends out to $\sim$3~arcmin.

\begin{figure}
\begin{center} 
\leavevmode 
\epsfxsize 0.95\hsize
\epsffile{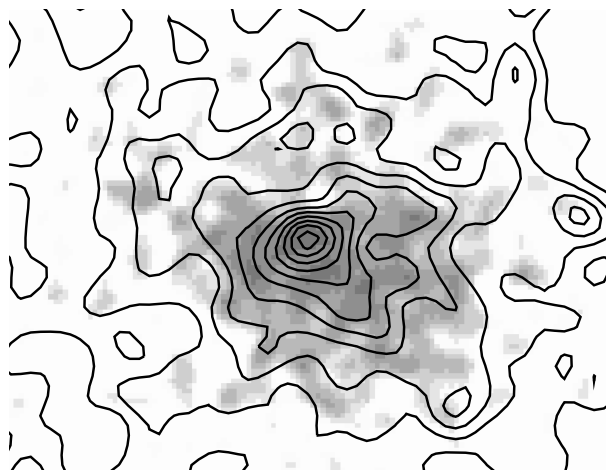}
\caption{X-ray contours of the adaptively smoothed mosaic \xmm image
in the (0.3-5.0)~keV energy range (as in Fig.~2) overlaid onto the
softness ratio image (S-H/S+H), where S is the soft band and H the
hard band [S=(0.3-1.0)~keV and H=(1.0-2.0)~keV)]. Darker shades of grey mark regions of softer emission. }\label{HR}
\end{center} 
\end{figure}

Generally, a galactic wake is expected to have softer emission than
the surrounding cluster. In order to investigate whether this is true
for the 4C34.16 wake, we constructed the softness ratio image shown in
Fig.~\ref{HR}.  Images for each \xmm camera were created from the
clean and filtered event lists in the (0.3-1.0)~keV, and (1.0-2.0)~keV
energy ranges. As before, we generated exposure-corrected mosaic
images in each energy band, using the {\sc SAS} task {\sc
emosaic}. The softness-ratio image of Fig.~\ref{HR} was calculated by
$\frac{S-H}{S+H}$, where $S$ is the softer and $H$ is the harder image
respectively. In order to better locate the emission that appears in
this image and to guide the eye, we overlay the contours of the
(0.3-5.0)~keV mosaic, presented in Fig.~\ref{contours}.  It is
apparent that the softest regions are dominated by emission from the
wake.

Thus, the above discussion shows that there is enhanced X-ray emission
from the region behind the galaxy, and that this emission is softer
than the surrounding cluster. 

\subsection{Wake's spectrum}\label{wake_spec}

Driven by the above results, we used the spectra of the wake to derive
its temperature and number density, to assist the investigations of
the wake production mechanisms that we will carry out later in this
paper.

In order to obtain the spectrum of the wake we followed the steps set
out in Section~\ref{cluster_spec}. The source and background
regions that were used are shown in Fig.~\ref{contours}.  The
background region was chosen so that the emission from the cluster at
the same off-source radius as the wake region is subtracted, while
avoiding regions close to the wake that might be contaminated by wake
emission.

The source spectrum was fitted in {\sc xspec} by an absorbed {\it
mekal} model, in a similar manner as done for the cluster in
Section~\ref{cluster_spec}. The hydrogen column ($N_{\rm H,w}$),
temperature ($kT$), and normalization ($norm$) were left as free
parameters. The metal abundances ($Z$) were held fixed to the value
found for the surrounding ICM ($Z$=0.25 ~${\rm Z_{\odot}}$,
Table~1). Given that we argue later in this paper that the wake
consists mainly of galactic material we fixed the metal abundances to
the higher value of $Z$=1 ~${\rm Z_{\odot}}$, and performed the same
fitting procedure. We obtained a sighly higher temperature for the
wake than the one found for $Z$=0.25 ~${\rm Z_{\odot}}$ (see Table 1)
of $1.44[1.02-2.19]$~keV. However, since there is not a well
established value for the metal abundance of central cluster galaxies,
and since the data do not allow us to measure it directly, we have no
reason to adopt a higher value for $Z$.

The spectral fit for the wake resulted in a normalization for the {\it
mekal} component of $norm=0.6[0.3-8.0] \times 10^{-18} \ {\rm
cm^{-5}}$. Using the dependency of $norm$ on the density ($n$) ($norm
\propto \int n^{2} \ dV$), we derived a density for the plasma in the
wake ($n_{\rm w}$) of $2.37[1.66-8.57] \times 10^{-3} \ {\rm cm
^{-3}}$. The volume of the wake ($V_{\rm w} = 1.6 \times 10^{70} \
{\rm cm^{3}}$) was calculated from the dimensions of the wake's region
shown in Fig.~2, assuming cylindrical symmetry (i.e., the wake is as
extended along the line-of-sight as on the plane of the sky), and that
it is on the plane of the sky. For the calculation of the wake's
density, we further assumed that the hot X-ray emitting component at
the derived temperature is distributed uniformly in the entire volume
of the wake.

However, the numerical simulations (Stevens et al. 1999; Acreman et
al. 2003) show that the average volume filling factor ($f$) for the
material that is at a temperature of $\sim$1~keV is between 0.3 and
0.4. For a given X-ray luminosity from a given volume, the inferred
density scales as $n_{\rm w}\propto f^{-0.5}$. Thus, it can be larger
than the calculated value of Table~1. For a filling factor of 0.35,
for example, the density of the non-uniformly distributed wake
material is 1.7 times larger than the tabulated value, bringing
it up to $4.03[2.82-14.57] \times 10^{-3} \ {\rm cm ^{-3}}$. Both
values are consistent within the errors.  In the following sections we
use the wake density given in Table~1, and a volume filling factor of
$f=1$, and we discuss, when necessary, the implications for $f<1$.

The wake's properties derived from the above spectral modelling are
presented in Table~1. The comparison of $N_{\rm H,w}$ with that found
from the cluster fits and the $N_{\rm H,G}$ indicates the presence of
some extra absorbing material ($N_{\rm H,w}-N_{\rm H,G} =
5.27[1.26-11.18] \times 10^{21} \ {\rm cm^{-2}}$) associated with the
wake. This excess of absorption towards the central parts of the
the Z0357.9+9432 was noted before in the \rosat data, and now with
\xmm we know that it is restricted to the wake region. However,
lacking the appropriate data (e.g., HI 21cm radio maps) we cannot be
confident that it is due to the presence of neutral material in the
wake. This neutral material could be stripped from the galaxy, 
and cause the absorption of the X-rays towards the wake.

\section{Discussion}\label{discussion}

\subsection{The bent radio galaxy 4C~34.16}\label{wat}

4C~34.16 is a wide-angle tailed (WAT) radio galaxy, and therefore shows an
abrupt `flaring' in its radio jets (Hardcastle \& Sakelliou 2004,
Jetha et al. 2005, Hardcastle, Sakelliou \& Worrall 2005). The jets in
these sources stop suddenly at some tens of kpc from the radio
core. After their termination point, large plumes emerge that are
thought to be shaped by the interactions with the surrounding medium.  

The jets in 4C~34.16 are transformed to plumes at $\sim$50~kpc from
the radio core. These plumes are bent symmetrically into a C-shape
(see Fig.~1). The main force acting on the plumes seems likely to be
the ram pressure resulting from the motion of the galaxy relative to
the ICM. The same motion should be responsible for the creation of the
wake, as the wake direction coincides with the direction in which the
plumes are bent.  Buoyancy makes an additional contribution, dragging
the plumes towards the outskirts of the cluster.  After the
termination point, the plumes can be traced out to a drojected
distance of $\sim$(120-140)~kpc.

By considering the balance of buoyancy and ram pressure forces
Sakelliou et al. (1996) used the {\it ROSAT} data to set an upper
limit on the galaxy velocity of $v_{\rm gal} < 300\ {\rm km \
s^{-1}}$.  With the new parameters derived from the \xmm observations
[$kT_{\rm ICM}=3.17$~keV compared to 1~keV used by Sakelliou et
al. (1996)] the allowed limits allowed by eq. (12) of Sakelliou et
al. (1996)  are somewhat higher. Depending on the exact orientation and
geometry of the flow inside the plumes, the required $v_{\rm gal}$
would be $ \sim 1000\ {\rm km \ s^{-1}}$ if the plumes are light
compared to the surrounding medium.  The projected length of the
plumes, on the other hand, sets a lower limit of $v_{\rm gal} > 1200 \
{\rm km \ s^{-1}}$ if we think that the plumes are just passive clouds
left behind by the motion of the galaxy. But we have independent
reasons to believe that the plume physics is more complicated, and we
cannot make accurate estimates of the galaxy speed from the radio
galaxy structure alone because we cannot measure the particle density
or fluid flow speeds in the plumes directly.

Taking into account the above discussion about $v_{\rm gal}$, in the
next section, we assess the wake production mechanisms demonstrating
the implications of a galaxy speed  
$> 1200 \ {\rm km \ s^{-1}}$.

\subsection{Wake production}\label{production}

As mentioned in the introduction, two physical processes can produce
galactic wakes: Bondi-Hoyle accretion and ram pressure stripping.  Of
course, both processes take place simultaneously (e.g., Acreman et
al. 2003). However, Bondi-Hoyle accretion generates pronounced wakes
when the galaxy velocity is sub- or trans-sonic, and their length does
not exceed a few kpc. For example, in a 3-keV cluster a Bondi-Hoyle
wake cannot be longer than $\sim$5~kpc (Sakelliou 2000). Additionally,
the accretion radius [$R_{\rm acc}=2 \ G \ M_{\rm gal} \ / \ (v_{\rm
gal}^{2}+c_s^{2})]$ is smaller in richer clusters; in a 3-keV cluster,
$R_{\rm acc}$ is $\sim$13~kpc if the galaxy is moving at the sound
speed, and it reaches only $\sim$19~kpc if $v_{\rm gal} \simeq 0.6
c_s$. Thus, in 4C~34.16 Bondi-Hoyle accretion can only supply the
galaxy with ICM material, increasing the mass of its ISM. The mass
accretion rate (\.{M}$_{\rm acc,BH}$) should depend on the galaxy and
cluster properties as \.{M}$_{\rm acc,BH} \propto \pi \ R_{\rm
acc}^{2} \ n_{\rm ICM} \ v_{\rm gal}$. Using a $v_{\rm gal}=890 \ {\rm
km \ s^{-1}}$ equal to the local sound speed ($c_s$), $n_{\rm
ICM}= \ {\rm 1.33 \times 10^{-3} \ cm^{-3}}$ (Table~1), and $R_{\rm
acc} \sim 13 \ {\rm kpc}$ (Sakelliou 2000), we find that the rate at
which ICM is accreted onto the galaxy is \.{M}$_{\rm acc,BH} \simeq 10
\ {\rm M_{\sun} \ yr^{-1}}$. It will be apparent that such accretion
rates are small compared to the mass of the wake that has already been
accumulated.

On the other hand, ram-pressure stripping dominates in the supersonic
regime. Taking into account the discussion of Section~\ref{wat} we can
deduce that the most possible production mechanism for the wake in
4C~34.16 is ram pressure stripping. Of course, some small contribution
(indirectly) from the ICM into the wake's body is also expected, but
it should not be substantial compared to the stripped material.

\subsubsection{The mass of the wake}

In Section~4.1 we measured the density of the wake (Table~1). 
Multiplying the density of the wake $n_{\rm w}$ by its volume $V_{\rm
w}$, we find a mass of $M_{\rm w}=2.3[1.6-8.4] \times 10^{10}
M_{\sun}$, which is comparable to the masses of the X-ray halos in
elliptical galaxies (e.g., Canizares, Fabbiano \& Trinchieri
1987). The material with $M_{\rm w}$ is contained in a region between
45 and 135~arcsec from the galactic core. As mentioned in
Section~\ref{wake_spec}, the filling factor of the cool wake material
may be smaller than 1. A filling factor of $f=0.35$ implies that the
mass of the wake is $f^{1/2}=0.59$ times the value calculated above,
but it is still within the quoted errors. A filling factor as small as
0.01 would be required to reduce the calculated mass by an order of
magnitude, but such a filling factor is unrealistically small.

If we assume that all the mass calculated above ($M_{\rm
w}=2.3[1.6-8.4] \times 10^{10} M_{\sun}$) has been accumulated during
the time taken for the galaxy to cross the current length of the wake
($\sim$130~kpc), we find that the wake must have been created during
the past $1.0 \times 10^{8}$~yr, where a galaxy velocity of $v_{\rm
gal}>1200 \ {\rm km \ s^{-1}}$ (Section~5.1) has been used. To create
such a wake over this time period a relatively large mass accretion
rate of \.{M}$_{\rm w} \simeq 200 \ {\rm M_{\sun} \ yr^{-1}}$ is
required. Of course, a volume filling factor of 0.35 would reduce
\.{M}$_{\rm w}$ to approximately half the quoted value, and only a
$f=0.01$ would reduce it by an order of magnitude, which would make it
comparable to \.{M}$_{\rm acc,BH}$ (see Section 5.2).  We will return
to the issue of this relatively large \.{M}$_{\rm w}$ in the following
sections.

\subsubsection{Comparison with the simulations}\label{simulations}

All the previous results and discussion suggest that the most probable
origin of the wake is through ram pressure stripping, and that the
galactic motion is trans- or mildly super-sonic. Additionally, from
Section~\ref{production} we have a first indication that the mass
accretion rate into the wake is substantial, and that the mass of the
material that has already accumulated is comparable to the total mass of the
X-ray halos in elliptical galaxies.

To test the above ideas, and to gain a better understanding under what
conditions such wakes can be created, we decided to compare the \xmm
results with hydrodynamical simulations of a galaxy moving through the
ICM of a cluster. Similar simulations have been presented elsewhere
(Stevens et al. 1999; Acreman et al. 2003). In our simulations, a
galaxy with mass of $M_{\rm gal}=2 \times 10^{12} \ M_{\sun}$ is set
into motion in a cluster of temperature $T_{\rm ICM}$ equal to the
value given in Table~\ref{tab1}.

\begin{table}
\begin{center}
 \caption{Simulations}\label{tsimul}
\begin{tabular}{cccc}	\hline \hline

(I)		&
(II)		&
(III)		&
(IV)		\\

\hline

No			&
$v_{\rm gal}$		&
$\alpha_{*}$		&
{\it \.{M}}$_{\rm rep}$	
\\

			&
(${\rm km \ s^{-1}}$)	&
($\alpha_{*,m}$)	&
($M_{\sun} \ {\rm yr^{-1}}$)	
\\

\hline

1	&
1200	&
27	&
92.1	
\\

2	&
1200	&
81	&
276.2	
\\

3	&
5000	&
27	&
92.1	
\\

4	&
5000	&
81	&
276.2	
\\

 \hline
\end{tabular}
\\
\vspace{0.2cm}
\begin{minipage}{8cm}
\small NOTES: $\alpha_{*,m}$ is the specific mass loss rate at the
present day [$\alpha_{*,m}=5.4 \times 10^{-20}~{\rm s^{-1}}$ (Mathews
1989)]. The mass {\it \.{M}} is calculated by {\it \.{M}}$_{\rm
rep}$=$\alpha_{*} \times M_{\rm gal}$(see \ref{simulations} for more
details).  \\
\end{minipage}
\end{center}
\end{table}

\begin{figure*}
\begin{center} 
\leavevmode 
\epsfxsize 0.45\hsize
\epsffile{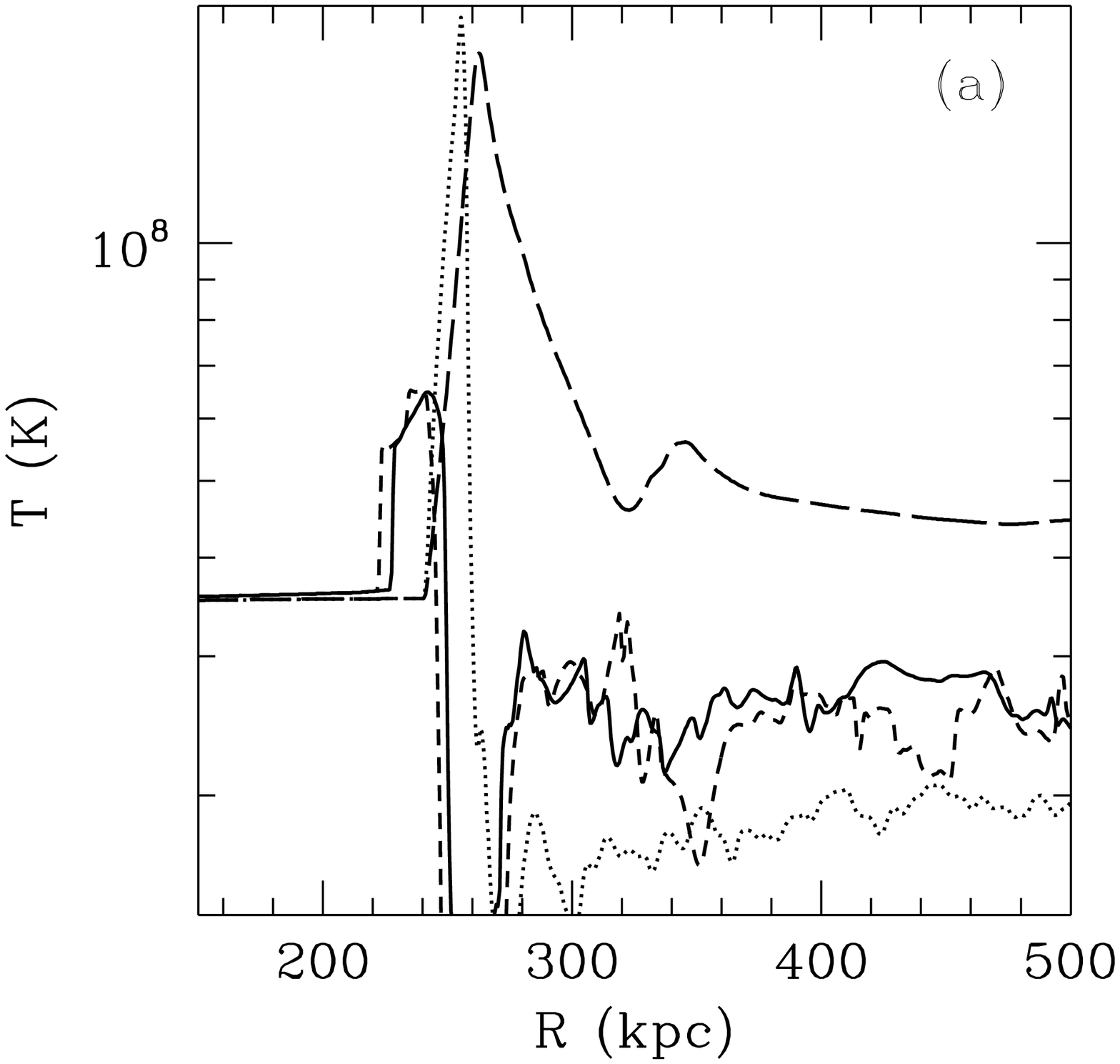}
\epsfxsize 0.45\hsize
\epsffile{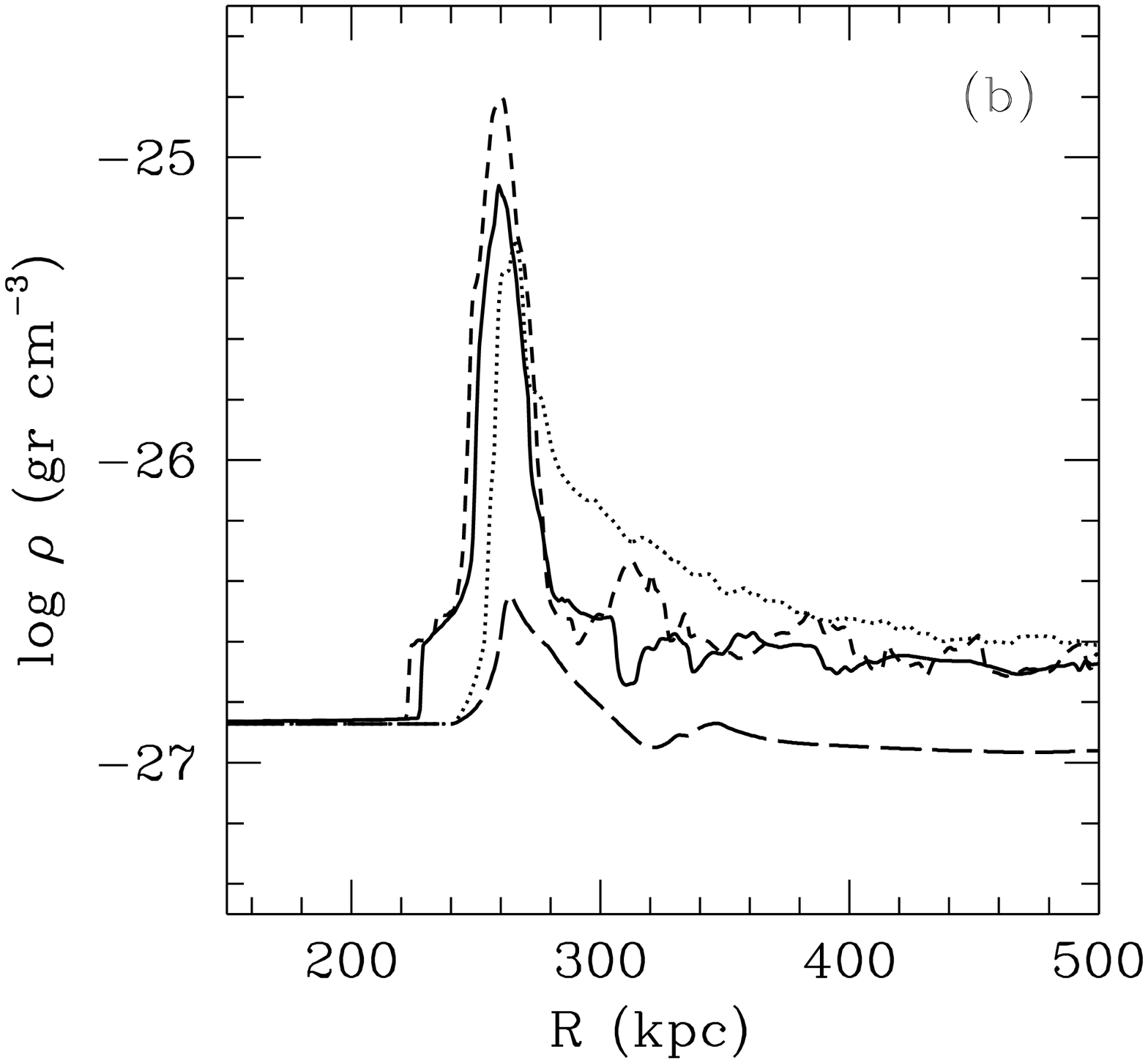}
\caption{Temperature (a) and density (b) variations along a line
parallel to the galaxy velocity. Both cuts are derived from the
simulated data. The galaxy is moving to the left of the image, and its
centre is at $\sim$260~kpc.  Simulation runs No 1, 2, 3, and 4, are
represented by a solid, short dashed, long dashed, and dotted lines
respectively (see Table~\ref{tsimul} for details).}\label{fsimul}
\end{center} 
\end{figure*}

We ran a number of simulations varying the galaxy velocity ($v_{\rm
gal}$) and specific mass loss ($\alpha_{*}$). We present here a
set of four runs that best show the dependency of the results on the
input parameters, and provide the best match to the \xmm results. The $v_{\rm
gal}$ and $\alpha_{*}$ used for each run are given in
Table~\ref{tsimul} in columns (II) and (III) respectively.  The
specific mass loss $\alpha_{*}$ is a measure of the mass that is
replenished ({\it \.{M}}$_{\rm rep}$) within the galaxy from a
combination of stellar mass loss and supernovae (see for more details
in Stevens et al. 1999). In Table~\ref{tsimul} [column (IV)] we also
list the {\it \.{M}}$_{\rm rep}$ used in the different simulation
runs, that is calculated by {\it \.{M}}$_{\rm rep}$=$\alpha_{*}
\times M_{\rm gal}$.

Temperature and density cuts along a line parallel to the direction of
the galactic motion are presented in Fig.~\ref{fsimul}(a) and (b)
respectively, using the simulated data from all four runs.  In
Fig.~\ref{fsimul}(a) we see that in the low velocity regime the
temperature of the wake does not reach the measured value, but is
on average $>$2.2~keV. However, increasing the mass replenishment rate
(run 2) results in the appearance of cold tongues of material that
emanate episodically from the galaxy. These cold features could bring
the temperature of the wake down to the measured values. Increasing
the mass replenishment rate means that more cold gas is available
within the galaxy to be stripped and added into the wake. However, in
simulation 2, the ram pressure is not strong enough to strip
the galaxy, and most of the cold gas ends up accumulating around the
galactic centre, increasing the density in the galactic core as seen
in Fig.~\ref{fsimul}(b). Higher galaxy velocities are required to make
this cold material accrete into the wake [Fig.~\ref{fsimul}(a)]. As
we see in the run 4, the temperature of the wake in the region
(320-460)~kpc (which is the distance from the radio core where we
measure the X-ray properties of the wake of Table~1) drops down to
$\sim$1.5~keV, which is within the range of wake temperatures we
measure.  In Fig.~\ref{fsimul}(b) we see that in the same region the
wake is overdense by a factor between 1.7 and 3.4,
again consistent with the measured ratio from the \xmm data (see Table~1).

Thus, a galaxy that is moving at a constant velocity $>1200 \ {\rm km
\ s^{-1}}$ through a medium with a uniform density equal to the
central density of the 4C~34.16 cluster, could produce a wake with the
measured properties, but {\it only if \.{M}$_{\rm rep}$ is as high as
$\sim300 \ M_{\sun} \ {\rm yr^{-1}}$}. Such mass replenishment rates
are high, especially for a central cluster galaxy which is a radio
galaxy. This becomes clearer if one thinks that the mass replenishment
rate in the nearby starburst galaxy M82 does not exceed a few tens of
$M_{\sun} \ {\rm yr^{-1}}$. Thus, we consider the previously presented
scenario unrealistic, as the host galaxy needs to contain a huge
amount of cold material, to be stripped by ram pressure and accreted
onto the wake.

\subsubsection{A more realistic scenario}

We have found that the mass of the wake is a substantial fraction of
the X-ray halo of an elliptical galaxy and large accretion rates are
required to produce it. The only alternative to the
Section's~\ref{simulations} scenario for the wake production is that
of a galaxy with a pre-existing halo falling into a cluster: a galaxy
with a substantial hot interstellar medium (ISM) has fallen into the
cluster, and is currently crossing the central region of the
cluster; its halo is stripped severely during this crossing, and most
of the pre-existing halo forms its wake.

The large-scale structure of the ICM provide support to this
picture. The \rosat images (Sakelliou et al. 1996) showed that the
ICM at large radii is elongated along a direction coincident with
the direction of the wake and the bent jets. This characteristic was
attributed to a recent disruption of the cluster by a small in-falling
group. Such a process has been demonstrated numerically with the recent
simulations of Acreman et al. (2003). They modelled a spherical galaxy
that is falling into a cluster not very different from the one around
4C~34.16 (they used a cluster temperature of 2.7~keV, while the ICM of
the 4C~34.16 cluster is at $\sim$3.2~keV (see
Section~\ref{cluster_spec}). Their work showed that: i) the galaxy
velocity becomes mildly super-sonic during the core passage (see their
fig.~1); ii) most of the halo is stripped during the first core
crossing (see fig.~3 and 4 in Acreman et al.); iii) the galaxy can lose
half or more of its ISM during its initial travel towards the cluster
centre.

According to the above scenario 4C~34.16 should not retain the large
quantities of ISM found in similar sources (e.g., Jetha et al. 2005,
Hardcastle et al. 2005). In order to see if this is true, we attempted
to model the galaxy spectrum in {\sc xspec}. The source region was a
circle of 10~arcsec radius around the active core. We fitted the
spectrum with a {\it mekal+power law} model modified by the Galactic
absorption. The temperature ($kT_{\rm ISM}$), power law index
($\Gamma$), and normalizations of the two models were left free to
vary. This fitting procedure resulted in a $kT_{\rm ISM}$ of 0.2~keV
and $\Gamma=2.3$ ($\chi^2/d.o.f$=15/14). However, the small number of
observed counts meant that the best-fitting parameters were not well
constrained. We found that the unabsorbed luminosity of the thermal
component only in the (0.6-3.5)~keV energy range is $\sim0.3 \times
10^{42} \ {\rm erg \ s^{-1}}$, lower than the luminosity of the ISM in
3C\,465 ($2 \times 10^{42} \ {\rm erg \ s^{-1}}$, Hardcastle et
al. 2005), suggesting that there might be a lack of hot gas from the
galaxy in 4C~34.16.

The close match of the Acreman et al. (2003) results and the X-ray
properties of the 4C~34.16 supports strongly the idea that the process
simulated by Acreman et al. (2003) represents the best explanation in
hand for what is happening around 4C~34.16.

\subsubsection{Is there a compression region in front of the galaxy?}

\begin{figure*}
\begin{center} 
\leavevmode 
\epsfxsize 0.45\hsize
\epsffile{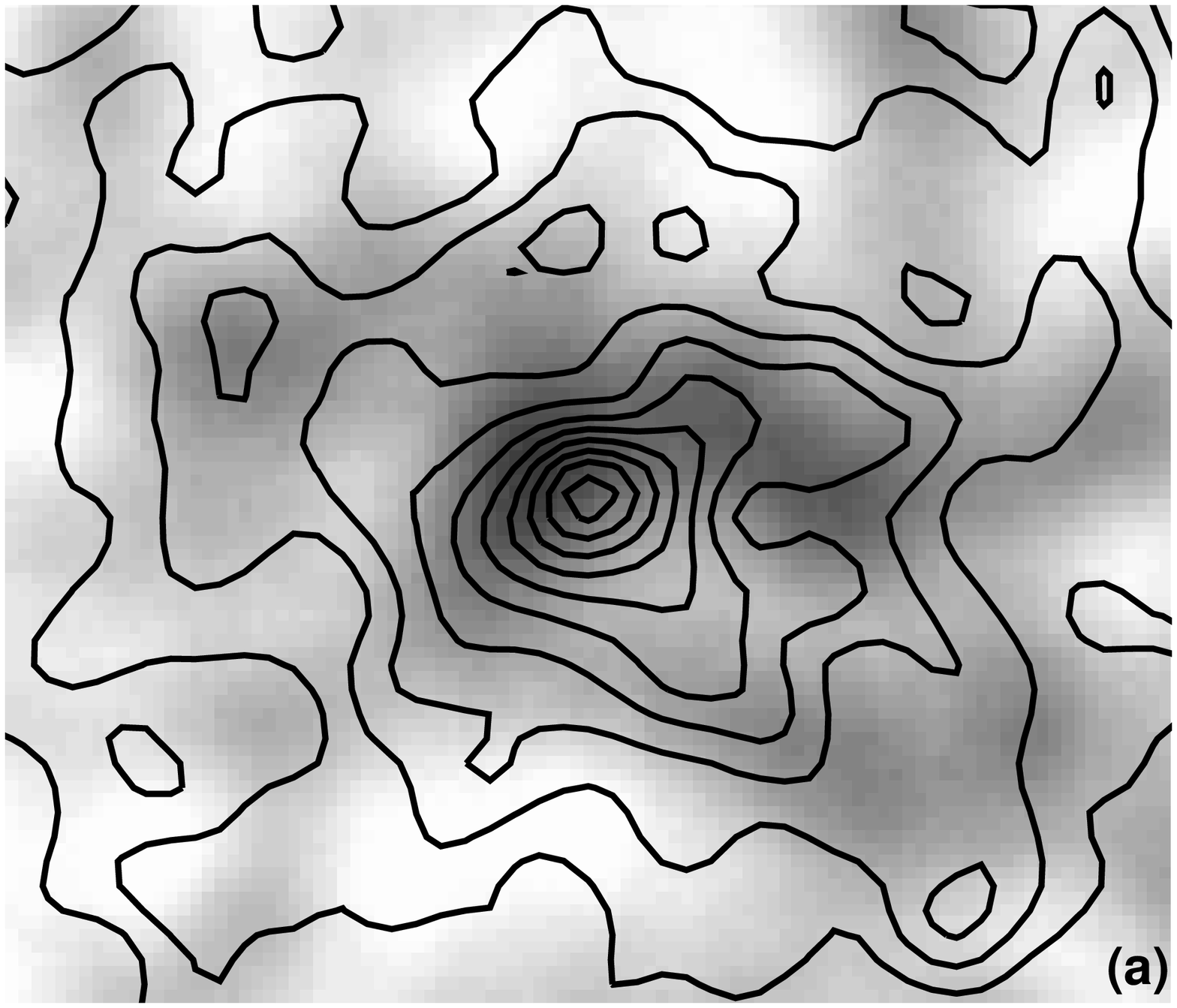}
\leavevmode 
\epsfxsize 0.45\hsize
\epsffile{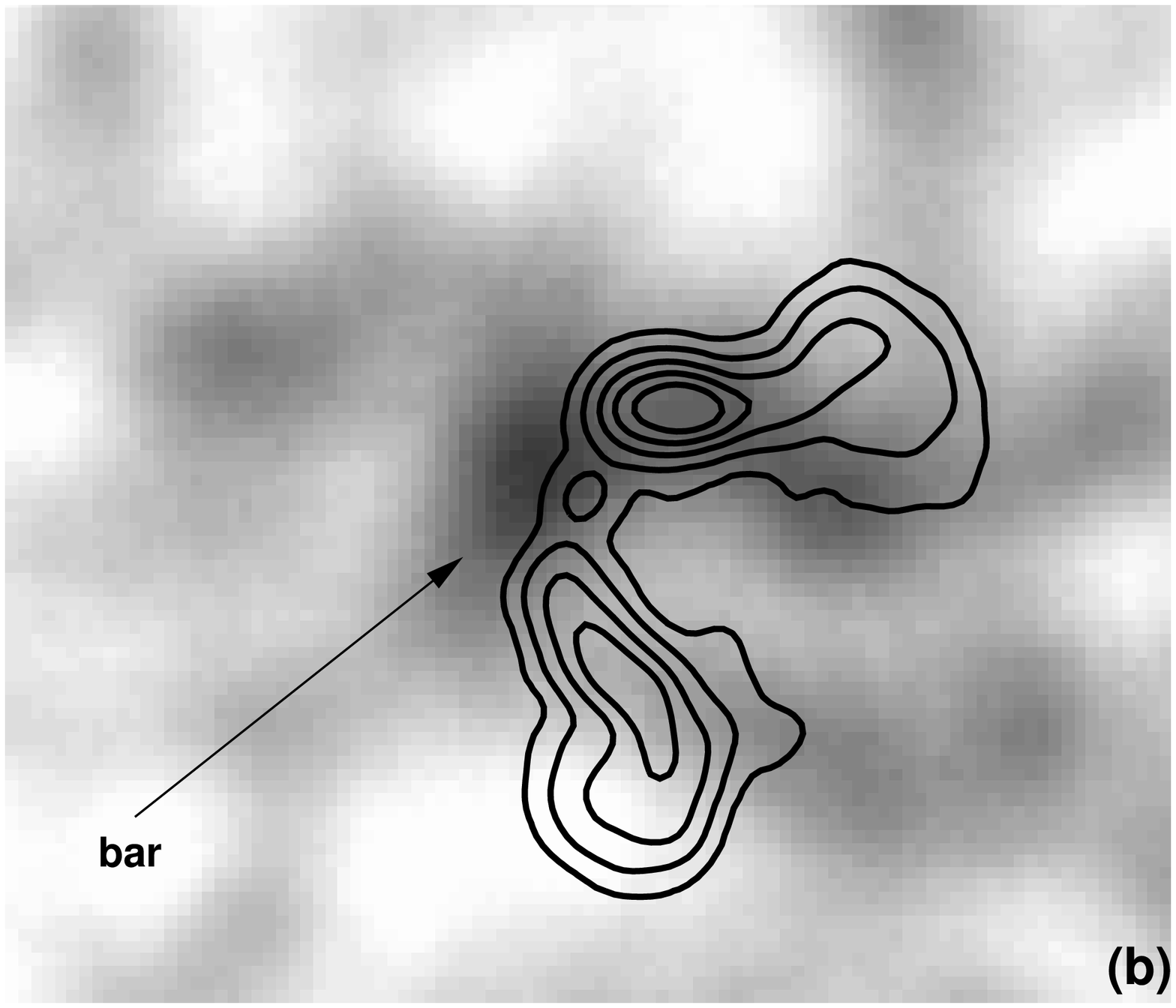}
\caption{(5.0-8.0)~keV image overlaid by (a) the (0.3-5.0)~keV
contours, and (b) the radio map.}\label{hard}
\end{center} 
\end{figure*}

According to the proposed scenario the host galaxy is currently
 crossing the cluster core and its velocity is mildly supersonic.
 Such a motion should compress the region in front of the
 galaxy, resulting in the heating of the ICM. The galaxy is moving relative to its surroundings  with a Mach number of $M >
 v_{\rm gal}/c_{\rm s} \sim 1200/890 = 1.35$, the compression factor
 is 1/$x > 1.5$. 
From Landau \& Lifshitz (1987) 
we find that such a compression would increase the temperature of the
 medium in front of the moving galaxy from $\sim$3.17~keV up to
 $kT_{\rm b}>$5~keV (in the above calculation we use $\gamma =
 5/3$). This effect can been seen clearly in the results of the
 simulations in Fig.~\ref{fsimul}(a), where the temperature in front
 of the galaxy is enhanced.

The presence of such a hot feature in the \xmm data would provide
further support for a supersonic motion.  Figure~\ref{hard} shows an
\xmm image in the (5.0-8.0)~keV range. Again, this mosaic is produced
in a similar manner to the ones of Section~2, and
smoothed with a Gaussian kernel with $\sigma=20$~arcsec. As can be
seen in this image the emission from the wake is not visible, as
expected, since we have found its emission to be soft and dominant in
the soft energy bands. The hard image reveals a potentially intriguing
feature. There appears to be a `bar' of bright emission in front of the
galaxy.  This region might be the above-mentioned compression region
in front of the galaxy.

The centroid of this `bar' is $\sim$18~arcsec
away from the galactic centre. One might think that this emission
comes from the radio core, but it is very unlike since at other
energy ranges there is a very good alignment between the radio core
and the X-ray peak. Additionally, we are not aware of pointing errors
as large as the measured offset between the `bar' and the radio
core. The `bar' appears extended and not point-like. It is detected
with a S/N$\sim$5 above the local background, which was calculated
from the same hard image between 1.5 and 5~arcmin away from the
galaxy. The inspection of optical images does not reveal any obvious
source that could be responsible for this hard X-ray emission. 

To obtain the spectrum of the `bar', we accumulated counts in an elliptical  region in-front of the galaxy with minor and major axes of 12 and 30~arcsec respectively. The background spectrum was taken in an annulus between 1 and
2.5 arcmin, similar to the one shown in Fig.~\ref{contours} . We fitted the spectrum with an absorbed {\it mekal}
model. The $N_{\rm H}$ and metal abundance were fixed to the Galactic
value and 0.25 respectively. Unfortunately, there are not enough
counts in the data to constrain the temperature of the detected
feature. The fits indicate that the temperature is $kT_{\rm b}>4$~keV
($\chi^{2}$/d.o.f.=27.3/27). Its unabsorbed luminosity is $L_{\rm
x}[(0.5-5.0)~{\rm keV}] \simeq 0.7 \times 10^{41} \ {\rm erg \
s^{-1}}$.

Applying the jump conditions from Landau \& Lifshitz (1987) for a Mach number of $M > 1.35$, we derived that 
the density of the `bar' should be $n_{\rm b}> 2.0 \times
10^{-3}\ {\rm cm^{-3}}$. Using {\sc pimms}, we find that the
luminosity of a thermal plasma with $T_{\rm b}$
and $n_{\rm b}$ properties derived above in the (0.5-5.0)~keV energy range is $>0.2
\times 10^{41}\ {\rm erg \ s^{-1}}$, consistent with the measured
value from the \xmm data. 

Thus, the properties of the hot feature in front of the galaxy might support
 the previous arguments that the motion of 4C~34.16 is currently 
supersonic as it is crossing the cluster core. However, better quality data are required to explore this possibility further.

\subsection{The fate of the wake}

Using the densities and temperatures of the ICM and the wake shown in
Table~1 we find that the wake is in pressure imbalance, by $P_{\rm w}/P_{\rm
ICM} \ \simeq \ 0.64$ (where $P_{\rm w}$ and $P_{\rm ICM}$ are the
thermal pressures of the wake and and ICM at the location of the wake,
respectively).
The pressure imbalance within the wake will be eliminated within
$t_{\rm P} \sim R_{\rm w} / c_s$, where $c_s$ is the local speed of
sound, and $R_{\rm w}$ is the radius of the wake. The size of the
overpressured wake, as measured from the X-ray data is (60 $\times$
120)~arcsec. Using an average radius of $R_{\rm w}$=65~kpc, we find
that the pressure imbalance will disappear in $t_{\rm P} \sim 0.7
\times 10^{8} \ {\rm yr}$, assuming that this region of wake material
is left behind in the cluster, and that it is not replenished by
neither stripped ISM nor accreted ICM. This time is slightly smaller
than the core crossing time ($t_{\rm S} \sim 2.3 \times 10^{8} \ {\rm
yr}$).

On the other hand, the entropy of the wake is less than the entropy of
the surrounding medium : $S_{\rm w}/S_{\rm ICM} \ \simeq \ 0.24$
(where $S_{\rm w}$ is the entropy of the wake and $S_{\rm ICM}$ the
entropy of the ICM at the location of the wake). Here, the entropy is
defined as $S =\frac{T}{n^{2/3}}$.  Entropy imbalances even out by the
transport of material towards regions in the surrounding cluster where
the entropies match. Thus, the low entropy wake will tend to go
towards lower-entropy regions in the cluster, which are at, or close to,
the cluster centre, which in the case of 4C~34.16, coincides with its host galaxy. Therefore, if a cool and dense wake is left
behind it will move towards the cluster centre by the buoyancy force:
\begin{equation}
\nabla F_{\rm B} \ = \ (\rho_{\rm w} - \rho_{\rm ICM}) \ g
\end{equation}
where $\rho_{\rm w}$ and $\rho_{\rm ICM}$ are the densities of the wake
and the ICM respectively, and $g$ is the cluster's gravitational
acceleration:
\begin{equation}
g = - \frac{kT_{\rm ICM}}{\mu \ m_{\rm p}} \ \frac{\nabla \rho_{\rm
ICM}}{\rho_{\rm ICM}}
\end{equation}
Thus, the wake will be driven by an acceleration of
\begin{equation}
\alpha = \left( 1 - \frac{\rho_{\rm ICM}}{\rho_{\rm w}} \right) \ g,
\end{equation}
where for the calculation of $g$ we used eq.~(2), and the properties of the cluster, as derived in earlier sections. 
The time that it will take for the wake to move to the cluster centre is
$t_{\rm E}^{2}=2r_{\rm w}/\alpha$, where $r_{\rm w}$ is the current
location of the wake relative to the cluster centre. For $r_{\rm
w} \simeq$1~arcmin, $t_{\rm E} = 3.3 \times 10^{8} \ {\rm yr}$.

The large uncertainties of the density of the wake (see Section~4.1) make the comparison of $t_{\rm P}$ and $t_{\rm E}$ difficult. With the current determination of the wake's properties it is unclear whether the
pressure imbalance will disappear quickly, leaving the entropy
structure almost unaltered. 

\section{Summary and conclusion}

We have presented \xmm observations of the X-ray wake that trails  
behind the WAT radio source 4C~34.16.

Our results can be summarized as follows:

\begin{itemize}

	\item{The wake is cooler and denser than the surrounding ICM: its temperature is $\sim$1.14~keV; for a filling factot of $f$=1 its density is 
	$\sim 2.6 \times 10^{-3} \ {\rm cm^{-3}}$, and for a more realistic $f$=0.35 it can be 1.7 times the above value.}

	\item{A comparison with hydrodynamical simulations shows that the
\xmm results fit well with the scenario simulated by Acreman et
al. (2003): the host galaxy of 4C~34.16 is falling into its cluster,
and currently it is crossing the core region, possibly moving
supersonically.}

	\item{As the galaxy passes through the cluster, it sweeps up
the ICM, forming a hotter and denser region in front of it. There is
evidence in the \xmm for the presence of such a feature, although
better quality data are needed to constrain its properties. }
	
\end{itemize}

\section*{Acknowledgments}

We thank Andrew Read for the use of his software and advice on \xmm
analysis, Chris Lee for interesting comments and his help with the
figures, and the von Hoerner \& Sulger GmbH for their kind
hospitality. We would also like to thank the anonymous referee for
useful comments that improved this paper. The Digitized Sky Survey,
and the NASA/IPAC Extragalactic Database have been used. The present
work is based on observations obtained with \xmm, an ESA science
mission with instruments and contributions directly funded by ESA
Member States and the USA (NASA). The National Radio Astronomy
Observatory (NRAO) is a facility of the National Science Foundation,
operated under a cooperative agreement by Associated Universities,
Inc.

\end{document}